\documentclass[11pt,a4paper,useAMS,usenatbib]{emulateapj}
\usepackage{epsfig}
\usepackage{amsmath}
\usepackage{natbib}
\usepackage{rotating}

\begin{document}

\title{Probing the hot X-ray gas \\ in the narrow-line region of Mrk~3}

\author{\'Akos Bogd\'an, Ralph P. Kraft, Daniel A. Evans, Felipe Andrade-Santos, and William R. Forman} 

\affil{Harvard-Smithsonian Center for Astrophysics, 60 Garden Street, Cambridge, MA 02138, USA}

\email{E-mail: abogdan@cfa.harvard.edu}

\shorttitle{Hot X-ray gas in the NLR of Mrk~3}
\shortauthors{BOGD\'AN ET AL.}

\begin{abstract}
We study the prototypical Seyfert 2 galaxy, Markarian 3, based on imaging and high-resolution spectroscopy observations taken by the \textit{Chandra} X-ray Observatory. We construct a deconvolved X-ray image, which reveals the S-shaped morphology of the hot gas in the narrow line region (NLR). While this morphology is similar to the radio and [O~III] emission, the distribution of the X-ray gas is broader than that obtained at these other wavelengths. By mapping the density and temperature distribution of the hot gas in the NLR, we demonstrate the presence of shocks towards the west ($M=2.5^{+1.0}_{-0.6}$) and east ($M=1.5^{+1.0}_{-0.5}$). Moreover, we compute the flux ratios between the  [O~III] and $0.5-2$~keV band X-ray luminosity and show that it is non-uniform in the NLR with the western side of the NLR being more highly ionized. In addition, based on the \textit{Chandra} grating data we investigate the line ratios of the Si XIII triplet, which are not consistent with pure photoionization. Based on these results, we suggest that in the NLR of Mrk~3 \textit{both} photoionization and collisional ionization act as excitation mechanisms.  We conclude that the canonical picture, in which photoionization is solely responsible for exciting the interstellar medium in the NLR of Seyfert galaxies, may be overly simplistic. Given that weak and small-scale radio jets are commonly detected in Seyfert galaxies, it is possible that shock heating plays a non-negligible role in the NLR of these galaxies. 
\end{abstract}


\section{Introduction}
\label{sec:introduction}
Large optical surveys demonstrated that galaxies evolve through mergers from star-forming spirals, through a transition region, to massive elliptical galaxies \citep[e.g.][]{bell04,faber07,schawinski14}. Outflows and the energetic feedback from active galactic nuclei (AGN) are widely believed to play a crucial role in building the observed luminosity function of galaxies and in the co-evolution of supermassive black holes and their host galaxies \citep[e.g.][]{benson03,croton06}. However, it is still debated how the energy released from AGN is coupled with the surrounding matter.  

The NLR, located beyond the sphere of influence of supermassive black holes, provides an ideal laboratory to explore the connection between the central AGN and the host galaxy. Given that the typical extent of NLRs is in the range of few hundreds to about a thousand pc, for nearby AGN these regions are well resolved, allowing detailed morphological and diagnostic studies. Investigations of the NLR reveal the presence of bright narrow emission lines at a wide range of energies, from [O~III] to X-rays \citep[e.g.][]{heckman14,netzer15}. The presence of emission lines at various wavelengths suggest a common ionization mechanism. A long-standing debate is whether the NLR of AGN is ionized by the nucleus or by shocks driven by radio jets. Although observational studies suggest that in radio galaxies jets may be responsible for the ionization of optical emission-line material \citep[e.g.][]{nesvadba08,lanz15}, the consensus is still lacking \citep[e.g.][]{robinson00}. Similarly, the dominant emission mechanism is also debated in radio-quiet AGN. Studies of nearby Seyfert galaxies suggest that their X-ray spectrum is consistent with photoionized gas  \citep{bianchi06}. The main arguments hinting that photoionization is the main excitation mechanism are (1) the morphological similarity between the diffuse X-ray  and the [O~III] emission; (2) the approximately constant flux ratios between the  [O~III] and soft X-ray emission; and (3) the acceptable fit obtained by describing the observed spectra with a photoionized gas models. However, several studies hint that the role of collisional ionization may be non-negligible in Seyfert galaxies \citep{capetti99,kukula99,maksym16}. To probe the ionization mechanism of the hot gas in the NLR of Seyfert galaxies, it is indispensable to perform detailed studies of nearby Seyferts with prime multi-wavelength data. As we demonstrate below, Mrk~3 is the ideal candidate for such a study. 

Mrk~3 (UGC~3426) is an early-type (S0) galaxy\footnote{Morphological classification taken from HyperLeda.} at $z=0.013509$, which hosts a luminous Seyfert 2 AGN. Given its brightness and proximity, Mrk~3 was the subject of a wide range of multi-wavelength observations. The \textit{Hubble Space Telescope} (\textit{HST}) [O~III] survey of nearby AGN showed that Mrk~3 is the second brightest source, after the spiral galaxy, NGC 1068 \citep{schmitt03}. The \textit{HST} images point out that the NLR of Mrk~3 exhibit a series of emission-line knots, which show an S-shaped morphology. Radio observations reveal the presence of a pair of jet knots, whose position angle is consistent with that of the NLR \citep{kukula93,kukula99}. Based on the spectroscopic study carried out with \textit{HST}, \citet{capetti99} argue that the NLR is a high-density shell that was shock heated by the jet. However, based on long-slit spectra obtained with \textit{HST} and by utilizing photoionization models, \citet{collins05,collins09} suggest that the NLR is dominated by photoionization. Thus, the mechanism responsible for ionizing the diffuse gas in the NLR of Mrk~3 remains a matter of debate. 

Mrk~3 has been explored with the X-ray grating spectrometers of \textit{Chandra} and \textit{XMM-Newton}. Based on the analysis of a 100 ks \textit{Chandra} HETG observation and by extracting the spectrum of an 8 pixel ($\approx4\arcsec$ wide) region,  \citet{sako00} suggested that the main excitation mechanism of the X-ray emitting plasma is photoionization. In agreement with this, the \textit{XMM-Newton} RGS spectrum of Mrk~3 also hinted that the soft X-ray emission is dominated by photoionized gas \citep{bianchi05}. However, previous  X-ray studies did not explore the NLR of Mrk~3 at spatial scales comparable to that of the \textit{HST} and radio images. Indeed, these works probed the entire NLR as a single region and did not take into account its complex structure.

\begin{table}
\caption{The list of analyzed \textit{Chandra} observations.}
\begin{minipage}{8.75cm}
\renewcommand{\arraystretch}{1.3}
\centering
\begin{tabular}{c c c c c }
\hline 
Obs ID &  $T_{\rm{obs}}$ (ks) & Grating & Date\\
\hline
873 & 100.6 & HETG & 2000/03/18 \\
12874 & 77.1 & HETG & 2011/04/19 \\
12875 & 29.9 & HETG & 2011/04/25 \\
13254 & 31.5 & HETG  & 2011/08/26 \\
13261 & 22.1 & HETG & 2011/05/02 \\
13263 & 19.7 & HETG & 2011/04/28  \\
13264 & 35.8 & HETG & 2011/04/27  \\
13406 & 21.4 & HETG & 2011/05/03  \\
14331 & 51.2 & HETG & 2011/08/28 \\ 
12293 & 30.6 & NONE & 2012/01/09 \\ \hline
\end{tabular} 
\end{minipage}
\label{tab:list1}
\end{table}

 \begin{figure*}
  \begin{center}
    \leavevmode
      \epsfxsize=16cm\epsfbox{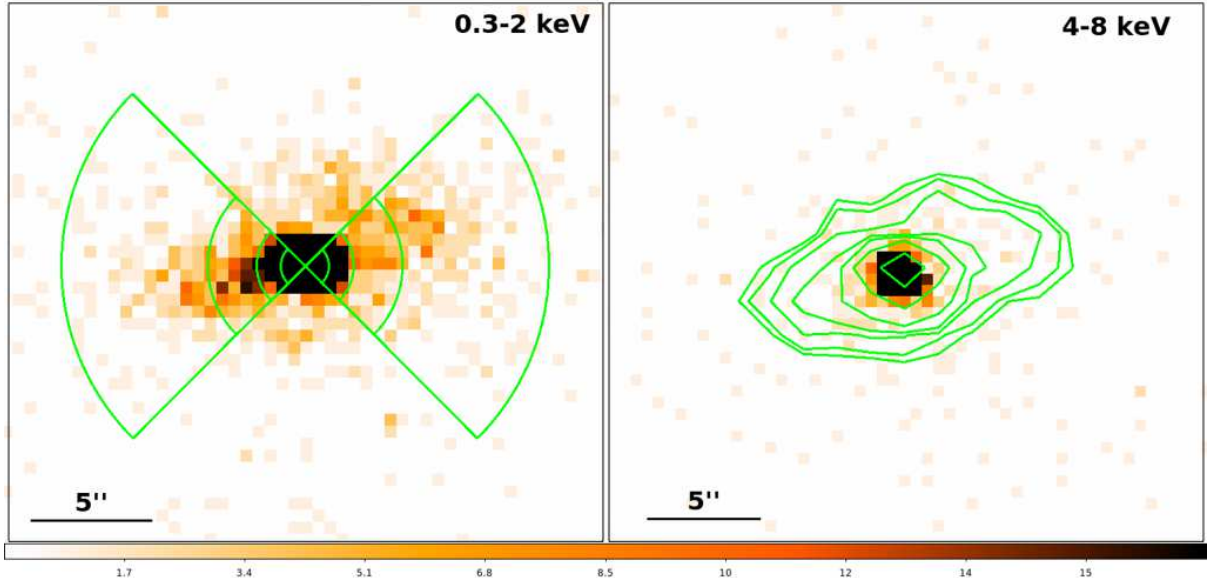}
      \caption{The raw \textit{Chandra} ACIS-S image of the central regions of Mrk~3 in the $0.3-2$ keV (soft) and $4-8$ keV (hard) band. The soft band image is extended and elongated in the east-west direction, whereas the hard band image appears to be point like. Based on the \textit{Chandra} PSF in the  $4-8$ keV band, about 90\% of the counts are expected to be enclosed within $2\arcsec$ radius for a point source. The hard band image demonstrates that more than $90\%$ of the counts are within this region, confirming that this image is consistent with a point source. In the $0.3-2$ keV band the 90\% enclosed fraction of counts corresponds to about $0.8\arcsec$. Only 35\% of the counts are within this region, demonstrating that besides the AGN another extended X-ray emitting component is present originating from ionized hot gas in the NLR of Mrk~3. In the soft band image we over plot the regions that were used to extract X-ray energy spectra for constructing the temperature profile (see Section \ref{sec:shock} and Figure \ref{fig:sb}). In the hard band image we over plot the logarithmic intensity levels taken from the $0.3-2$ keV band image, which are in stark contrast with the point like distribution of the counts in the $4-8$ keV band.}
\vspace{0.5cm}
     \label{fig:zero}
  \end{center}
\end{figure*}

Given the prototypical nature of Mrk~3 and the wealth of available multi-wavelength data, it is a prime laboratory to explore the mechanisms responsible for exciting the diffuse X-ray gas. In this work, we focus on analyzing \textit{Chandra} observations of Mrk~3 to distinguish between collisional ionization from the small-scale radio jet and photoionization from the AGN radiation field. This project is greatly facilitated by the deep \textit{Chandra} HETG observations that allow us to perform high-resolution spectroscopy of the NLR at $0.5\arcsec$ spatial scales -- nearly an order of magnitude smaller scales than applied in previous X-ray grating studies.  

This work is structured as follows. In Section 2 we introduce the data and describe the main steps of the analysis. In Section 3 we present our results, namely discuss the deconvolved X-ray image of the diffuse emission, study the surface brightness and temperature structure of the hot gas, and probe the \textit{Chandra} HETG spectra in seven distinct locations as a function of radius from the nucleus. We discuss our results in Section 4 and argue that both photoionization and collisional ionization play a role in the NLR of Mrk~3. We summarize in Section 5.  
	
The luminosity distance of Mrk~3 is $D_{\rm L} = 58.8$ Mpc and the corresponding angular scale is $278 \ \rm{pc \ arcsec^{-1}}$. All uncertainties listed in the paper are $1\sigma$ errors.

 \begin{figure}[t]
  \begin{center}
    \leavevmode
      \epsfxsize=8.5cm\epsfbox{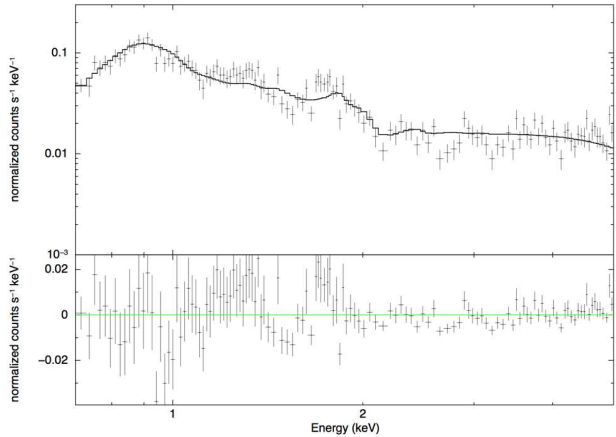}
      \caption{X-ray energy spectrum of a circular region with $2\arcsec$ radius centered on the nucleus of Mrk~3. The spectrum was fit with a two component model consisting of a thermal model and a power law model. The column density was fixed at the Galactic value. The best-fit model is over plotted. The bottom panel shows the residuals of the fit.}
\vspace{0.7cm}
     \label{fig:spec_central}
  \end{center}
\end{figure}

\section{The Chandra data}
\label{sec:chandra}
The \textit{Chandra} X-ray Observatory observed Mrk~3 in nine pointings with HETG/ACIS-S for a total of  389.3 ks. In addition, one pointing with an exposure time of 30.6 ks was done with \textit{Chandra} ACIS-S in imaging mode. The details of the individual observations are listed in Table \ref{tab:list1}. The data were reduced with standard CIAO\footnote{http://cxc.harvard.edu/ciao/} software package tools (CIAO version 4.5, CALDB version 4.6.7). 

To analyze the HETG data, we reprocessed all observations, which assures that the most recent calibration updates are applied. We used standard CIAO tools to create the region masks (\textit{tg\_create\_mask}) and extract the spectra (\textit{tg\_extract}). Throughout the analysis we only consider the first order dispersed spectra for the Medium Energy Grating (MEG) and High Energy Grating (HEG). To maximize the signal-to-noise ratios of the spectra, we combined the $\pm1$ orders of each grating. 

To probe the spectral variability of Mrk~3, we investigated the individual exposures and found that the spectra are consistent and the count rates measured in the $6-10 \ \rm{\AA}$ wavelength range exhibit $\lesssim7\%$ variations. Therefore, we combined the spectra from all individual observations to obtain a single first order HEG and MEG spectrum. Finally, we produced grating response files for each observations by employing the \textit{mkgarf} and \textit{mkgrmf} tools, which were then combined. 

The imaging observation was analyzed following the main steps outlined in \citet{emery17}. First, we used the \textit{chandra\_repro} tool to reprocess the observations. Then we searched for high background periods using a light curve that was extracted from the $2.3-7.3$ keV energy range, which band is the most sensitive to flares \citep{hickox06}. Using the \textit{deflare} tool and applying $3\sigma$ clipping, we did not find any high background time periods, hence the total exposure time of the imaging observation remains $30.6$ ks. Given that we aim to explore the gaseous X-ray emission around Mrk~3, bright point sources -- mostly originating from low-mass X-ray binaries or background AGN -- need to be identified and removed. To detect the point sources, we utilized the  \textit{wavdetect} tool. The resulting source regions were excluded from the analysis of the diffuse emission. Although we identify several point sources, including the nuclear source associated with the galaxy, aside from the central AGN none of the sources are in the proximity of the NLR. To account for the background emission when studying the diffuse emission, we utilize nearby regions, which assures the precise subtraction of both the instrumental and sky background components. Exposure maps were produced for the images to correct for vignetting effects. To create the exposure maps, we used a spectral weights file that was computed by assuming an optically-thin thermal plasma emission model with $N_{\rm H} = 9.67 \times 10^{20} \ \rm{cm^{-2}}$ column density, $kT= 0.85$ keV temperature, and $Z=0.24$ Solar metallicity. This model represents the best-fit average spectrum of the hot gaseous emission in the NLR of Mrk~3 (see Section \ref{sec:basic}).

 \begin{figure*}[t]
  \begin{center}
    \leavevmode
      \epsfxsize=5.75in\epsfbox{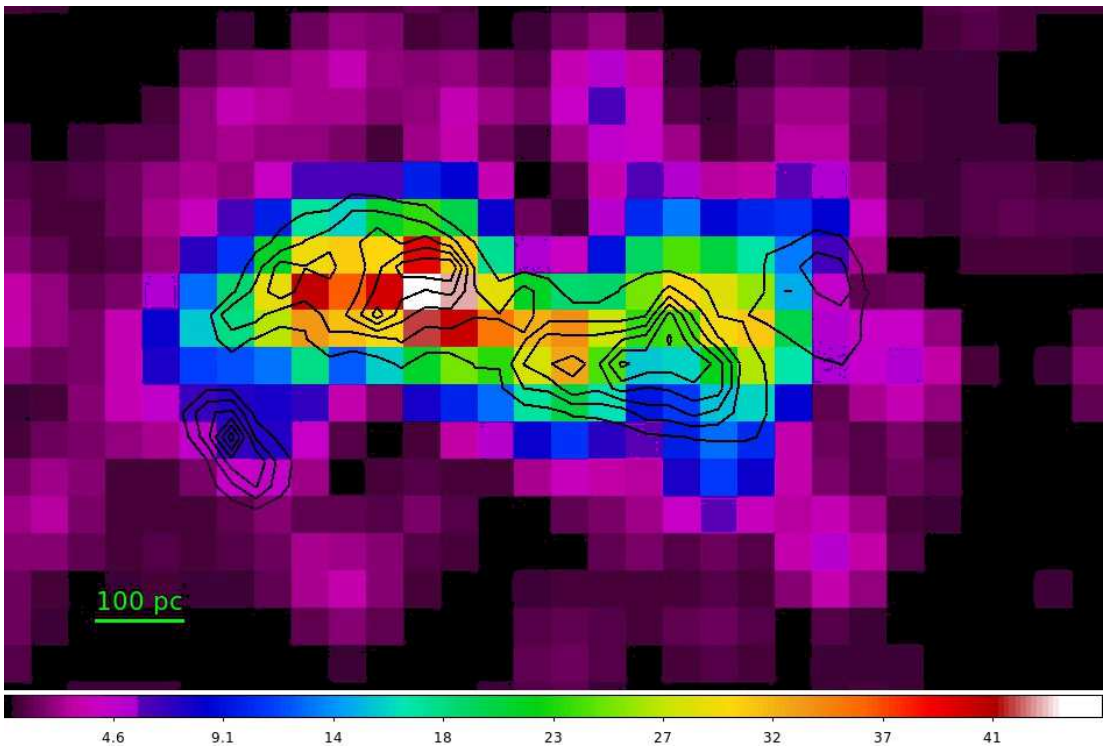}
      \caption{Deconvolved $0.3-2$ keV band \textit{Chandra} image of the NLR, where one pixel corresponds to $0.148\arcsec$ ($41.1$ pc). The scale bar corresponds to $\approx0.36\arcsec$. Over plotted are the [O~III] intensity levels taken from the \textit{HST} Faint Object Camera. The contour levels are $[1.67, 3.33, 5.00, 6.67, 8.33, 10.00]\times10^{-16} \ \rm{erg \ s^{-1} \ cm^{-2}}$ . The overall morphology of the X-ray and [O~III] images are similar, but the X-ray emission exhibits a broader distribution. The [O~III]-to-X-ray flux ratios are non-uniform across the east-west axis: we observed $\mathcal{R}_{\rm [O~III]/X} \approx 5.6 $ towards the east and $\mathcal{R}_{\rm [O~III]/X} \approx 2.2 $ towards the west. While these flux ratios are consistent with those obtained for other Seyfert galaxies, the higher ionization state towards the west indicates that shock heating may play a role.}
\vspace{0.7cm}
     \label{fig:deconvolveoiii}
  \end{center}
\end{figure*}

\section{Results}

\subsection{X-ray images of Mrk~3}
\label{sec:image}

In Figure~\ref{fig:zero} we depict the $0.3-2$ keV (soft) and $4-8$ keV (hard) band X-ray images of the central $20\arcsec \times 20\arcsec $ ($5.56 \times 5.56 $ kpc) region around Mrk~3 based on the sole imaging observation (Obs ID: 12293). Both images reveal the presence of a bright nuclear point source. However, the overall distribution of X-ray photons is strikingly different. The hard band image appears to be round and symmetric, whereas the soft band image shows an elongated structure in the east-west direction.

To probe whether the distribution of photons can be explained by a bright point source, we construct the \textit{Chandra} point spread function (PSF) for both energy ranges. Based on the hard  band PSF we expect that $\sim90\%$ of the photons should be enclosed within a circular region with radius of $2\arcsec$. In agreement with this, we find that somewhat more than $90\%$ of the photons are encircled within this radius, implying that the hard band emission can be explained with the bright AGN. The PSF extracted for the soft band predicts that the $90\%$ encircled radius is $0.8\arcsec$. However, within this radius only $\sim35\%$ of the photons are included, implying that beyond the nuclear source an extended X-ray emitting component is present. This diffuse X-ray emitting component in the NLR of Mrk~3, originating from hot X-ray gas, is in the main focus of our study.  

\subsection{Average properties of the hot gas in the NLR}
\label{sec:basic}

We establish the nature and average characteristics of the extended emission within the NLR by extracting an X-ray energy spectrum using the ACIS-S imaging observation. We utilize a circular region with $2\arcsec$ ($556$ pc) radius centered on the center of Mrk~3. We note that this region covers most of the NLR. We fit the resulting spectrum with a two component model consisting of an absorbed optically-thin thermal plasma emission model (\textsc{APEC}) and a power law model. The thermal component describes the gaseous emission, while the power law component accounts for the emission associated with the nuclear source and the population of unresolved X-ray binaries. The column density was fixed at the Galactic value. The spectrum and the best-fit model is shown in Figure \ref{fig:spec_central}. 

Based on the fit performed in the $0.5-2$ keV band, we confirm the presence of a significant gaseous component. The average best-fit temperature of the hot gas is $kT=0.83\pm0.03$ keV and the metallicity of $Z=0.24^{+0.24}_{-0.09}$ Solar using the \citet{anders89} abundance table. Given the stellar mass of the galaxy ($M_{\rm \star} = 1.6 \times 10^{11} \ \rm{M_{\odot}}$), the metallicity of the gas is relatively low. Indeed, other massive early-type galaxies exhibit approximately Solar metallicities \citep[e.g.][]{ji09}, whereas lower mass gas-poor ellipticals have sub-Solar metallicities \citep{bogdan12}, similar to that observed in Mrk~3. The slope of the power law component is $\Gamma = 1.96\pm0.10$, which is similar to that obtained by \citet{guainazzi16}, who performed a thorough analysis of the spectral properties of the AGN. In addition, these authors reported that Mrk 3 has a heavily absorbed continuum emission with $N_{\rm H} = (0.8-1.1)\times10^{24} \ \rm{cm^{-2}}$. However, due to the high absorbing column this emission component does not add a notable contribution at energies below $\lesssim5$ keV, hence our results are not affected by this emission in any significant way. 

The absorption corrected $0.3-2$ keV band luminosity of the thermal component is $L_{\rm 0.3-2keV} = 6.7\times10^{40} \ \rm{erg \ s^{-1}}$. We note that the observed X-ray luminosity and the gas temperature is in broad agreement with the scaling relations established for massive early-type galaxies \citep{goulding16}. Based on the best-fit spectral model we compute the emission measure of the gas and obtain $\int n_e n_H dV = 7.9\times10^{63} \ \rm{cm^{-3}} $. By using an admittedly simplistic approach and assuming uniform density and spherical symmetry for the gas distribution, we estimate the average gas density $n_e=0.61 \ \rm {cm^{-3}}$ and obtain a total gas mass of $M = 1.1\times10^7 \ \rm{M_{\odot}}$ within the studied volume. This gas mass is comparable to that obtained by \citet{collins09} from \textit{HST} observations of the NLR.

 \begin{figure*}[t]
  \begin{center}
    \leavevmode
      \epsfxsize=5.75in\epsfbox{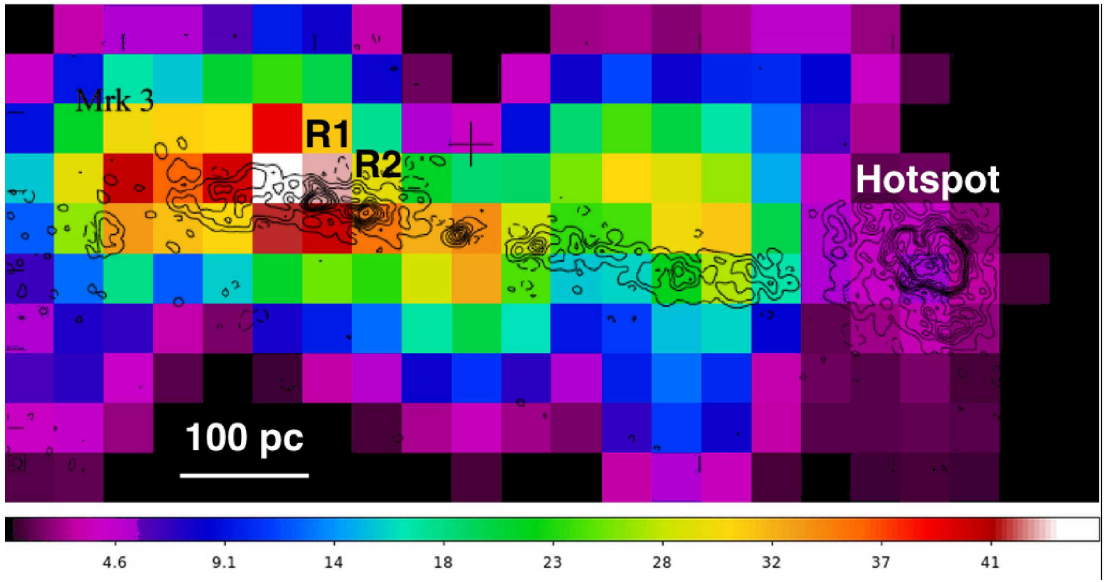}
      \caption{Same as Figure \ref{fig:deconvolveoiii}, but over plotted are the intensity levels of the 18 cm EVN and Merlin radio image \citep{kukula99}. The intensity levels of the contours are identical with those presented in Figure 2 of \citet{kukula99}. The image demonstrates the S-shaped morphology of the extended X-ray emission, which is in good agreement with the structure of the radio jet. Note that the X-ray emission envelopes the radio emission. The asymmetric radio jets, which extend to about $200$ pc and $400$ pc toward the east and west, signify the presence of a strong shock towards the west and a weaker shock towards the east  \citep{kukula99}. The main features in the radio emission, namely the hotspot on the western side and the two bright radio components on the eastern side, are highlighted and further discussed in Section \ref{sec:radio}. }
\vspace{0.7cm}
     \label{fig:deconvolve}
  \end{center}
\end{figure*}

\subsection{High-resolution images} 
\label{sec:deconvolved}
\subsubsection{Deconvolved X-ray image}

High-resolution radio and optical observations demonstrate the complex structure of the NLR. Given that the native $0.492\arcsec$ per pixel \textit{Chandra} resolution is lower than the resolution of the radio and optical images, for a more appropriate comparison we enhance the resolution of the ACIS imaging. This allows us to explore the spatial structure of the diffuse X-ray emission at finer angular scales. To this end, we apply the Lucy-Richardson deconvolution algorithm.  Since the observed \textit{Chandra} image is the intrinsic brightness distribution of the source (in our case the hot gas in the NLR of Mrk~3) convolved with the point spread function (PSF) of the detector, it is indispensable to have a good understanding of the PSF. To construct an accurate image of the PSF, we used the Chandra Ray Tracer (ChaRT). Specifically, we ran a ray-trace simulation using the ChaRT web interface\footnote{http://cxc.harvard.edu/ciao/PSFs/chart2/runchart.html}, which set of rays was then projected onto the detector-plane, resulting in a pseudo-event file. We binned this PSF event file to a fraction of the native ACIS pixels and created an image of the PSF in the $0.5-2$ keV band. 

For the Lucy-Richardson deconvolution we utilized the $0.5-2$ keV band X-ray image binned to $30\%$ of the native ACIS resolution and the similarly binned PSF image obtained from ChaRT. We used the \textsc{CIAO} \textit{arestore} task to carry out the deconvolution and iterated 100 times. While we tried to construct deconvolved images at different resolutions, we found that for the NLR of Mrk~3 the best result is achieved if $30\%$ of the ACIS resolution is applied. 

The deconvolved \textit{Chandra} image, shown in Figures~\ref{fig:deconvolveoiii} and \ref{fig:deconvolve}, have a pixel size of $0.148\arcsec$. We note that the applied Lucy-Richardson deconvolution technique tends to sharpen features. Therefore, the true X-ray light distribution is slightly more extended than seen on the deconvolved images.

\begin{figure*}[t]
  \begin{center}
    \leavevmode
      \epsfxsize=8.5cm\epsfbox{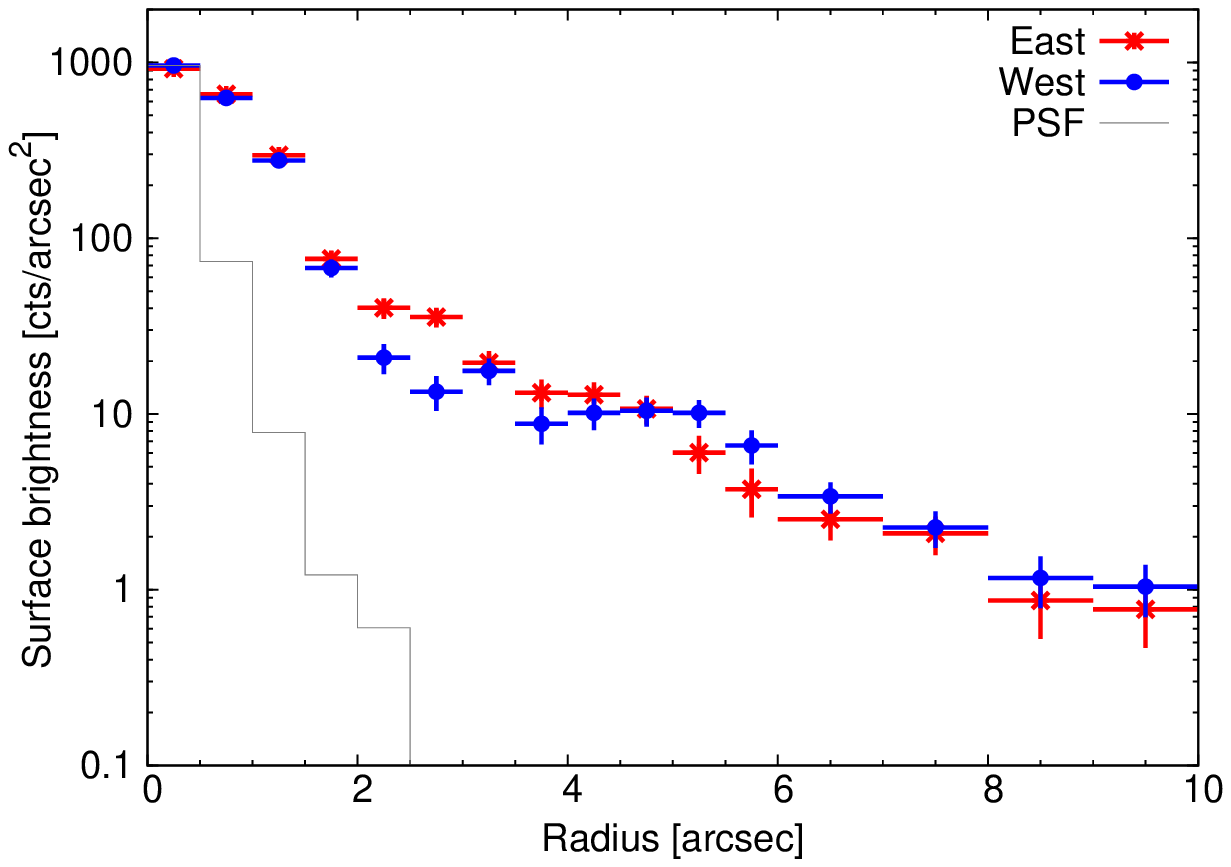}
      \hspace{0.5cm}
      \epsfxsize=8.5cm\epsfbox{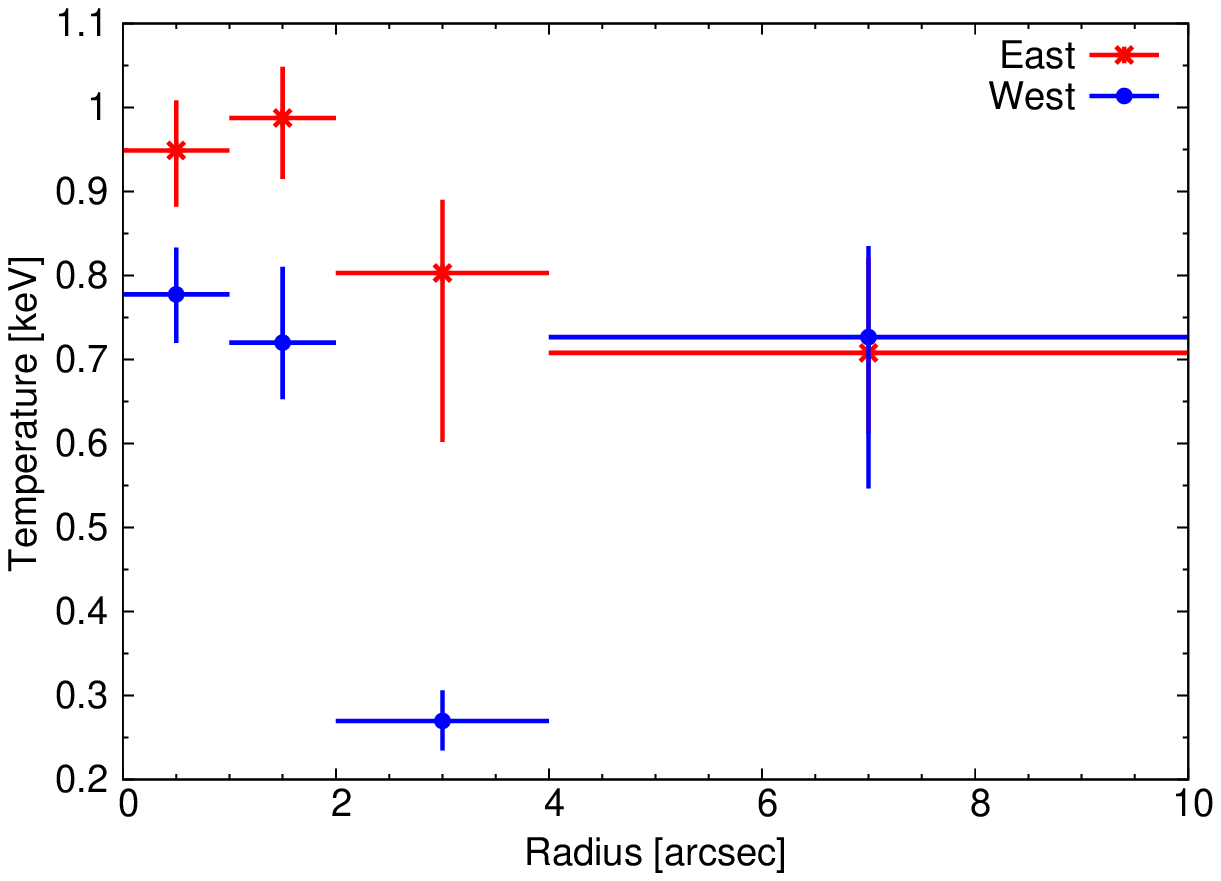}
\vspace{0.3cm}
      \caption{$0.3-2$ keV band X-ray surface brightness (left panel) and temperature (right panel) profiles of the diffuse gas towards the east and west of the NLR. On the left panel the solid histogram shows the PSF obtained from \textit{Chandra} ray tracing. Note the drop in surface brightness and temperature at $\sim2\arcsec$ towards the east and west. The observed surface brightness (hence density) and temperature ratios across the edge signify the presence of shock fronts. Based on the de-projected density jumps and the Rankine-Hugonoit jump conditions we derive shock Mach-numbers of $M=2.5^{+1.0}_{-0.6}$ and $M=1.5\pm0.2$ towards the west and east, respectively. The temperature jumps suggest comparable Mach numbers.}
\vspace{0.7cm}
     \label{fig:sb}
  \end{center}
\end{figure*}

\subsubsection{Comparing the X-ray and [O~III] morphology}
The morphological similarity and the nearly uniform flux ratios between the [O~III] line emission and the gaseous X-ray emission were used to argue that photoionization is the main excitation mechanism in Seyfert galaxies  \citep{bianchi06}. In addition, \citet{bianchi06} studied a sample of radio galaxies and obtained similar conclusions. To this end, we confront the [O~III] and X-ray morphology and flux ratios ($\mathcal{R}_{\rm [O~III]/X} = F_{\rm [O~III]}/F_{\rm 0.5-2keV}$) in the NLR of Mrk~3. In Figure~\ref{fig:deconvolveoiii} we present the deconvolved \textit{Chandra} image of the central regions of Mrk~3 and over plot the intensity levels of the [O~III] $\lambda5007$ emission observed by the \textit{HST}. The [O~III] image was taken with the Faint Object Camera at an angular resolution of $\approx0.1\arcsec$.  There is an overall agreement between the distribution of the X-ray light and the [O~III] intensity levels as both images exhibit a characteristic S-shaped morphology. However, the emission from the hot X-ray gas has a broader distribution and surrounds the [O~III] emission.

To compute the flux ratios, we utilize the [O~III] fluxes measured by \citet{collins05} and the X-ray luminosity obtained from the $0.5-2$ keV band \textit{Chandra} images. Given the  different angular resolution of the two images, we derive the average flux ratios in two regions corresponding to the east and west of the NLR within $1 \arcsec$ radius from the nucleus. The flux ratios are different on the two sides of the AGN. Specifically, we obtain $\mathcal{R}_{\rm [O~III]/X} \approx 5.6 $ towards the east and a significantly lower value, $\mathcal{R}_{\rm [O~III]/X} \approx 2.2 $, towards the west. The former ratio agrees with those obtained by \citet{bianchi06}, suggesting that photoionization plays a notable role to excite the gas.  However, the $\mathcal{R}_{\rm [O~III]/X}$ ratio on the western side of Mrk~3 is significantly lower and is comparable with sources that contain small-scale radio sources \citep{bianchi06,balmaverde12}. Therefore, the higher ionization state towards the west hints that the interaction between the jet and ISM may play -- at least -- a complementary role in the ionization of the gas. Thus, the non-uniform flux ratios and the broader distribution of the X-ray emission in the NLR of Mrk~3 suggest that photoionization may not be the sole excitation mechanism.  

\bigskip

\subsubsection{Comparing the X-ray and radio morphology}
\label{sec:radio}
The high-resolution 18 cm EVN and Merlin radio images of Mrk~3 (see the intensity levels in Figure \ref{fig:deconvolve}) reveal jets with an S-shaped structure and a remarkable hotspot on the western side \citep{kukula93,kukula99}. These authors suggest that the S-shaped morphology of the radio jet may either be due to a change in the jet axis or from the jet interacting with the rotating interstellar medium. Moreover, they suggest that the characteristics of the hotspot on the west may signify the presence of a shock, where the radio jet is interacting with the surrounding material. On the eastern side a similar hotspot is not observed, but two bright radio components are present at $<100$ pc from the nucleus. These features are marked as R1 and R2 in Figure \ref{fig:deconvolve}. \citet{kukula99} suggest that these radio components may have played a role in thermalizing the kinetic energy of the eastern jet,  hence reducing the jet's Mach number and leading to a weaker eastern shock. 

To compare the morphology of the X-ray and radio emission,  in Figure  \ref{fig:deconvolve} we show the  deconvolved X-ray image with the 18 cm radio intensity levels over plotted. This image reveals that the overall morphology between the X-ray and radio structure is similar since both images show the S-shaped morphology. However, the radio jets are narrow, whereas the gaseous X-ray emission is significantly broader and surrounds the radio emission. This hints that collisional ionization may play a role and the gas may be driven by shocks \citep{wilson01,massaro09}. Based on the prediction of shocks in the X-ray surface brightness, we investigate the X-ray data to identify potential shocks in the NLR.

\subsection{Detection of shocks in the NLR}
\label{sec:shock}

To search for possible shocks in the NLR, we investigate the surface brightness and temperature distribution of the X-ray gas. We extract the profiles using circular wedges with position angles of $135\degr-225\degr$ and $315\degr-405\degr$, where $0\degr$ and $90\degr$ correspond to west and north, respectively. While both the surface brightness and temperature profiles are extracted using wedges with these position angles, the width of the individual wedges are different. To extract the surface brightness profiles, we used regions with widths of $0.5\arcsec-1\arcsec$ and for the temperature profile the extraction regions had widths of $1\arcsec-6\arcsec$ depending on the brightness of the diffuse emission. 

The surface brightness profiles, extracted from the $0.3-2$ keV energy range towards the east and west of the NLR, are depicted in the left panel of Figure \ref{fig:sb}. Along with the surface brightness profiles, we also show the expected brightness distribution of the PSF obtained from ChaRT. As discussed in Section \ref{sec:image}, the PSF has a significantly narrower distribution than the diffuse emission, thereby demonstrating that the extended emission cannot be associated with the bright nuclear source. The surface brightness profiles reveal a notable jump at $\sim2\arcsec$ towards the east and west. 

\begin{figure*}
  \begin{center}
    \leavevmode
      \epsfxsize=18.5cm\epsfbox{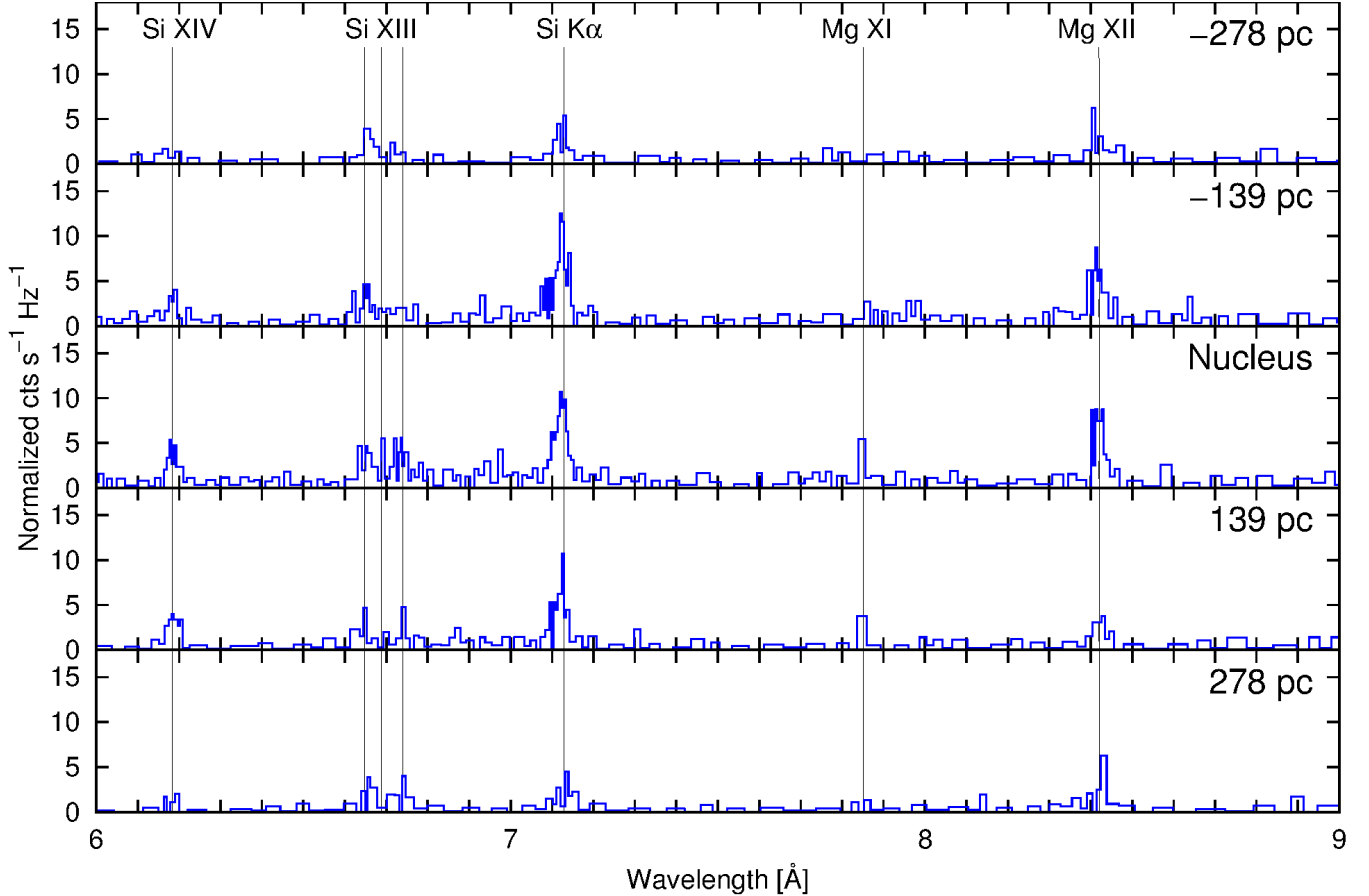}
      \vspace{-1.2cm}
      \caption{The \textit{Chandra} MEG spectra of the NLR of Mrk~3 in the $6-9 \ \rm{\AA}$ wavelength range. The five panels show five different extraction regions from the east (top panel) towards the west (bottom panel). The extraction regions had a width of $0.5\arcsec$ ($139$ pc). The systemic position of the major lines are marked with solid lines. The wavelengths are corrected for the cosmological redshift (z=0.013509) of Mrk~3. Note that we do not detect a significant redshift and blueshift between the east and west sides of the NLR or relative to the systemic position of the emission lines.}
     \label{fig:spec1}
  \end{center}
\end{figure*}

Although the surface brightness profile demonstrates the presence of jumps, de-projection analysis needs to be performed to determine the exact position and the magnitude of the corresponding density jumps. Therefore, we utilize the \textsc{proffit} software package tools \citep{eckert11} and construct de-projected density profiles. We assume spherical symmetry for the gas density within each wedge and assume that the gas density can be described with a broken power law model inside and outside the edge. We obtain density jumps of $n_1/n_0=2.68\pm0.54$ at $r_{\rm cut} = 2.14\arcsec \pm 0.08\arcsec$ (or $595\pm22$ pc) towards the west and $n_1/n_0=1.72\pm0.26$ at $r_{\rm cut} = 1.80\arcsec \pm0.08\arcsec$ (or $478\pm22$ pc) towards the east. 

In the right panel of Figure \ref{fig:sb}, we show the temperature profile of the hot gas towards the east and west. To measure the temperature of the gas, we construct X-ray spectra of each region and fit them with a model consisting of an \textsc{APEC} thermal emission model with the metallicities fixed at $Z=0.24$ Solar (Section \ref{sec:basic}) and a power law model. The latter component accounts for the emission arising from the AGN at $r\lesssim2\arcsec$ and from the population of unresolved low-mass X-ray binaries at $r\gtrsim 2 \arcsec$ \citep{irwin03,gilfanov04}. The best-fit spectra are depicted in the Appendix A. The profiles reveal a significant drop at $\sim2\arcsec$  towards the west  and a smaller jump towards the east. Specifically, we observe a temperature drop of $T_{1}/T_{0}=1.23\pm0.24$ and $T_{1}/T_{0}=2.67\pm0.39$ towards the east and west. Based on the presence of sharp surface brightness jump, and the observed density and temperature ratios across the edge, we conclude that the observed discontinuity on the western side of Mrk~3 is a shock front  \citep[e.g.][]{markevitch07,emery17}. 

To compute the shock Mach number ($v\equiv c_{\rm s}$) and the corresponding velocity, we utilize the Rankine-Hugoniot jump conditions \citep{landau59,markevitch07}, which directly connect the pre-shock and post-shock density and temperature with the Mach number. Given that we measure the density and temperature jumps in the NLR of Mrk~3, we can derive the Mach numbers using these two independent approaches. Based on the pre-shock and post-shock densities we find $M=2.5^{+1.0}_{-0.6}$ towards the west and $M=1.5\pm0.2$ towards the east. Based on the magnitude of the temperature jump, we derive the shock Mach numbers of  $M=2.4\pm0.3$ and $M=1.5^{+1.0}_{-0.5}$ towards the west and east, respectively. We emphasize that the Mach numbers obtained from the density and temperature jumps are in excellent agreement with each other. For a 0.7 keV plasma the sound speed is $c_{\rm s} = \sqrt{(\gamma kT)/( \mu m_{\rm H})} = 420 \ \rm{km \ s^{-1}}$ by using $\gamma = 5/3$ and $\mu = 0.62 $. Hence, the velocities are $v = 1050^{+420}_{-252} \ \rm{km \ s^{-1}}$ and $v = 630\pm84 \ \rm{km \ s^{-1}}$ towards the west and east, respectively.

The presence of shocks suggests that the hot gas in the NLR of Mrk~3 is undergoing collisional ionization due to the interaction between the radio jet and the circum nuclear material.

\subsection{High-resolution spectrum of the NLR}
\label{sec:ratios}

The bipolar morphology of the X-ray emitting gas combined with the presence of small-scale radio jets and the detection of shocks suggest the presence of an outflow.  To characterize the properties of the outflow, we utilize the HETG spectrum of the NLR. Due to the relative proximity of Mrk~3 and the superb angular resolution of \textit{Chandra}, we can perform spatially resolved X-ray spectral diagnostics on the outflow. To this end, we study the emission line spectra at two locations on either side of the nucleus with extraction regions that have a width of $0.5\arcsec$ ($139$ pc). Although the outflow is traced out to larger radii in the ACIS images, the signal-to-noise ratio is not sufficiently high to explore the high-resolution spectrum of the outflow beyond these regions. 

In Figure \ref{fig:spec1} we show the dispersed spectra of the individual extraction regions in the most relevant $6-9 \ \mathrm{\AA}$ region. To obtain these spectra, the plus and minus orders were combined, and the depicted wavelengths are corrected for the cosmological redshift. To fit the spectral lines, we utilize a model consisting of an absorbed power law model to account for the continuum and a series of Gaussian lines. When fitting the lines, we fixed the slope of the power law at $\Gamma=1.7$, but left the normalization as a free parameter.  Additionally, the centroid, width, and normalization of the Gaussian lines were also free parameters. The best fit line centroid wavelengths were corrected for the cosmological redshift and then compared with the laboratory measurements of strong lines based on the NIST data base \citep{verner96}. 

The spectra reveal a series of H-like and He-like emission lines along with fluorescence lines. In general, the set of identified lines and their best-fit wavelengths are in good agreement with those of \citet{sako00}. In this work, we compare the best-fit line centroids of the strongest emission lines between the east and west side of the nucleus and probe whether the outflowing gas shows significant blue-shift or redshift. Detailed modeling of the emission spectrum lines will be subject of a future paper. 

Based on the HETG spectra we find that, within measurement uncertainties, the line centroids towards the east and west agree with the laboratory wavelengths (Table \ref{tab:list2}). In addition, they do not show a statistically significant difference between the east and west side of the NLR. Specifically, all lines are within the $1\sigma$ uncertainties with the expected wavelengths, except for the Si line we measure a $1.5\sigma$ offset from the laboratory wavelength. In the absence of red-shifted and blue-shifted line centroids, we place upper limits on the outflow velocity of the hot gas. The upper limits typically remain below a few hundred $ \rm{km \ s^{-1}}$. The detailed constraints on the outflow velocity of the gas are listed in (Table \ref{tab:list3}). 

We note that these velocities are significantly lower than those inferred from the Rankine-Hugonoit jump conditions (Section \ref{sec:shock}). This difference is likely caused by the orientation of NLR, which has an inclination of $5\degr$ implying that is it is virtually in the plane of the sky \citep{crenshaw10}. Therefore, if the outflowing gas propagates along the plane of the sky and does not have a significant velocity towards (and away) from the observer, the projected velocities will be close to 0. Hence, the low outflow velocities computed from the HETG data indirectly point out that the outflow propagates almost in the plane of the sky. 

The low observed outflow velocities observed from the \textit{Chandra} HETG data are at odds with the results of \citet{capetti99}, who identified [O~III] emission lines shifted with several hundreds $\rm{km \ s^{-1}}$. We speculate that the observed velocity difference might be due to the decoupled nature of the cold [O~III] and the hot X-ray gas. In this picture, the rotating cold gas has a different velocity and temperature structure than the X-ray gas.  Therefore, the shock driven by the radio jet will not drag the cold and hot gas components with the same velocity, implying that these gaseous components remain decoupled. In addition, we mention that due to the $\approx0.1\arcsec$ angular resolution of the \textit{HST} Faint Object Camera, \citet{capetti99} extracted narrow regions that were mostly coincident with the locations of bright radio components. As opposed to these, the \textit{Chandra} HETG spectra cover notably larger regions with $0.5\arcsec$ width, implying that these regions include brighter and fainter parts of the emission. This difference might also contribute to the observed velocity difference. However, further exploring the velocity difference would require a dedicated analysis, which is beyond the scope of this paper.

\begin{figure}
  \begin{center}
    \leavevmode
      \epsfxsize=8.5cm\epsfbox{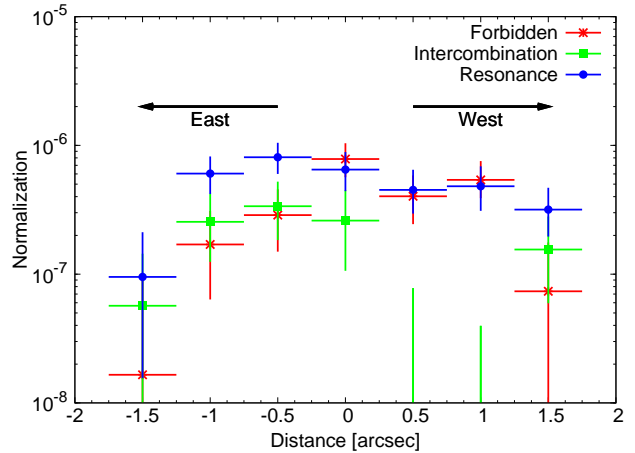}
      \caption{The intensities of the resonance, intercombination, and forbidden lines of the Si XIII triplet as a function of distance from the nucleus based on \textit{Chandra} HETG data. The east (negative distances) and west (positive distances) sides show different line ratios, despite the comparable $\mathcal{G}$-ratios. The line ratios hint that multiple ionization processes may play a role. In the central regions and towards the east the photo ionization is dominant, while collisional ionization may play a non-negligible role towards the west.}
\vspace{0.7cm}
     \label{fig:gratio}
  \end{center}
\end{figure}

\subsection{Line ratios of He-like ions}
\label{sec:ratios}
The line ratios of He-like triplets, and in particular the $G$ ratios, are suitable to probe the ionization state of the gas. Due to the high energy resolution of HETG, the three most intense lines, namely the resonance ($\rm{1s^2 \ ^1S_0 - 1s2p \ ^1P_1 }$), the intercombination ($\rm{1s^2 \ ^1S_0 - 1s2p \ ^3P_{2,1} }$), and the forbidden lines ($\rm{1s^2 \ ^1S_0 - 1s2s \ ^3S_1 }$), can be individually resolved. Following \citet{porquet10}, we derive the $G$ ratio as: 
$$ G (T_{\rm{e}}) = \frac{F+I}{R} \ .$$  
 where $R$, $I$, and $F$ refer to the resonance, intercombination, and forbidden line strengths, respectively. 

In Mrk~3, the most prominent He-like ion is the Si XIII triplet at $\sim6.7 \ \rm{\AA}$. The He-like lines of Mg and Ne are also detected, but these lines are significantly weaker due to the lower effective area of MEG and the relatively high absorbing column, and hence, cannot be used to compute constraining G ratios.

Based on a 100 ks \textit{Chandra} HETG observation \citet{sako00}  concluded that the G-ratios are inconsistent with a pure collisional plasma and are marginally consistent with a photoionized plasma. However, these line ratios were obtained by treating the entire NLR as a single region. Due to the presence of shocks, it is feasible that the line ratios show variation in the east-west direction. Therefore, we characterize the ionization state of the gas as a function of central distance by computing the $G$ ratios of the Si XIII triplet in seven distinct locations. The central region is centered on the nucleus of Mrk~3 and has a width of $0.5\arcsec$, while the regions at $0.5\arcsec$ (139 pc), $1\arcsec$ (278 pc), and $1.5\arcsec$ (417 pc) radii towards the east and west each comprise $0.5\arcsec$ wide extraction regions. 

To fit the lines, we used Gaussian line profiles (\textsc{agauss} in \textsc{XSpec}). Based on the fits we find that the G-ratios in the NLR are in the range of $\mathcal{G} = 0.7-1.1$ and $\mathcal{G} = 1.6\pm0.7$ in the center. Although the G-ratios are comparable within uncertainties at every radius, the line strengths of the resonance, intercombination, and forbidden lines exhibit stark differences. As demonstrated in Figure \ref{fig:gratio}, the intercombination line is weak or virtually absent towards the west, while it is prominent in the central region and in the east of the NLR. In addition, the intensities of the resonance and forbidden lines are comparable towards the west, while the resonance lines are factor of about $3$ times stronger than the forbidden lines towards the east. These results hint that multiple processes may be responsible for ionizing the hot gas.

\begin{table*}
\footnotesize
\centering
\caption{The list of strong emission lines observable in the nucleus and in the NLR of Mrk~3.}
\begin{minipage}{14cm}
\renewcommand{\arraystretch}{1.8}
\begin{tabular}{c c c c c c c}
\hline
 & & -278 pc$^\dagger$  & -139 pc$^\dagger$  & Nucleus & 139 pc$^\dagger$  & 278 pc$^\dagger$   \\
\hline
Ion & $\lambda_{\rm{lab}}$ & $\lambda_{\rm{obs}}$  & $\lambda_{\rm{obs}}$  & $\lambda_{\rm{obs}}$ & $\lambda_{\rm{obs}}$  & $\lambda_{\rm{obs}}$   \\
& ($\rm{\AA}$)  &($\rm{\AA}$) & ($\rm{\AA}$)  & ($\rm{\AA}$)  & ($\rm{\AA}$)  & ($\rm{\AA}$)   \\
\hline
Si XIV &  6.183 & $6.173^{+0.023}_{-0.019}$ & $6.177^{+0.012}_{-0.003}$ &  $6.184^{+0.005}_{-0.004}$ &  $6.185^{+0.007}_{-0.003}$ & $6.186^{+0.007}_{-0.003}$  \\

Si XIII (R)$^\ddagger$ & 6.648 & $6.661^{+0.010}_{-0.007}$  & $6.667^{+0.016}_{-0.021}$ & $6.656^{+0.011}_{-0.011}$   & $6.644^{+0.006}_{-0.010}$   & $6.644^{+0.005}_{-0.004}$   \\

Si XIII (F)$^\ddagger$ & 6.740 & $-$ & $-$ & $6.737^{+0.005}_{-0.008}$  & $6.741^{+0.022}_{-0.007}$  & $6.739^{+0.005}_{-0.013}$   \\

Si K$\alpha$ & 7.128 &  $7.122^{+0.009}_{-0.008}$  & $7.123^{+0.002}_{-0.002}$  & $7.121^{+0.002}_{-0.002}$ & $7.121^{+0.003}_{-0.002}$  & $7.121^{+0.003}_{-0.002}$   \\

Mg XII & 8.420 &  $8.408^{+0.012}_{-0.010}$ & $8.412^{+0.009}_{-0.003}$   & $8.421^{+0.003}_{-0.003}$  & $8.422^{+0.010}_{-0.009}$  &
 $8.421^{+0.012}_{-0.009}$  \\

Ne X & 12.132  & $12.115^{+0.011}_{-0.009}$ & $12.131^{+0.015}_{-0.011}$  & $12.132^{+0.011}_{-0.018}$  & $12.135^{+0.011}_{-0.012}$  & $12.135^{+0.010}_{-0.012}$  \\
 \hline
\end{tabular} 
\flushleft
$^\dagger$ Negative and positive offsets correspond to east and west, respectively. \\
$^\ddagger$(R) and (F) represent the resonance and forbidden lines, respectively.  \\
\end{minipage}
\label{tab:list2}	
\end{table*}

Although the G-ratios are similar to those expected for collisional plasma, the observed values may be influenced by resonance line scattering, which is relevant for high absorbing column densities ($N_{\rm HI} \gtrsim 10^{21} \ \rm{cm^{-2}} $). This, in turn, could enhance the intensities of the resonance lines, thereby decreasing the G-ratios of a photoionized plasma and mimicking collisional ionization \citep{porquet01,porquet10}. To probe whether resonance line scattering plays a role in the NLR of Mrk~3, we rely on \citet{collins05} who probed the geometry of the NLR and the extinction as a function of angular position. These authors found that Mrk~3 hosts inner gas disks, which results in a positive extinction gradient from west to east. Specifically, \citet{collins05} measured $E(B-V)=0.12-0.16$ towards the west and $E(B-V)=0.2-0.4$ towards the east. We convert these values to hydrogen column densities following \citet{shull85} as $N_{\rm HI}=5.2 \times 10^{21} \ \rm{cm^{-2}} \times \rm{E(B-V)}$, and conclude that the  $E(B-V)$ color excess corresponds to $N_{\rm HI} = (0.6-0.8) \times 10^{21} \ \rm{cm^{-2}}$ and $N_{\rm HI} = (1.0-2.1) \times 10^{21} \ \rm{cm^{-2}}$ towards the east and west, respectively. Thus, resonance line scattering is expected to increase the resonance line intensities, and hence decrease the G-ratios towards the west. As opposed to this, due to the relatively low column densities, resonance line scattering is not expected to significantly influence the observed G-ratios towards the east.

Overall, the line intensities of the Si XIII triplet  and the $G$-ratios hint that both excitation mechanisms -- photoionization and collisional ionization -- may be present in the NLR of Mrk~3. Specifically, in the central regions and towards the east the main ionizing mechanism may be photoionization, whereas collisional ionization may play a role on the west.

\section{Discussion}
\label{sec:discussion}
\subsection{Excitation mechanisms}
There is a significant debate about the ionization process of the thermal gas in the NLR of Seyfert galaxies. The observed X-ray emission may either originate from photoionized gas or may be due to gas shock heated by the radio jet. Detailed morphological studies of a sample of Seyfert galaxies pointed out the nearly constant OIII-to-X-ray flux ratios in the NLR \citep{bianchi06}. Specifically, these studies found the median value of $\mathcal{R}_{[O~III]/X} = 5 $ and a scatter of about 0.3 dex. These arguments suggest that a common ionizing source, photoionization from the nuclear source, may be responsible for the observed emission. However, this simple picture may break down when galaxies with small-scale radio jets are investigated. These galaxies exhibit lower O~III-to-X-ray flux ratios, indicating a higher level of ionization. This implies that photoionization may not be the only ionizing source, but the interaction between the radio jets and the dense ISM may also play a role. The picture, in which photoionization is the main ionization mechanism, is further challenged when the morphology of the X-ray gas and radio emission is compared. Specifically, in several radio galaxies (e.g. 3C 293, 3C 305, NGC 4258) the X-ray emission exhibits a broader distribution than the radio jets, hinting that shock heating may play a role in heating the gas to X-ray temperatures \citep[e.g.][]{wilson01,massaro09,lanz15}.
 
Our results obtained for the NLR of Mrk~3 can be summarized as follows. 
\begin{itemize}
\item The X-ray gas and the [O~III] emission share similar morphology. However, the X-ray light distribution is more extended in the east-west direction than the [O~III] emission. 
\item The [O~III]-to-X-ray flux ratios are non-uniform across the NLR. In the central regions and in the east they are $\mathcal{R}_{\rm [O~III]/X} \approx 5.6 $, while towards the west the observed ratio drops to  $\mathcal{R}_{\rm [O~III]/X} \approx 2.2 $.
\item The X-ray and radio morphology shows generally similar structures, but the X-ray emission is significantly broader and surrounds the radio emission. 
\item We detect shocks with $M= 2.4\pm0.3$ and $M=1.5^{+1.0}_{-0.5}$ toward the west and east, respectively.  The shock front towards the west is approximately consistent with the locations of the radio hot spot. 
\item The line ratios of the Si XIII triplets do not favor photoionization as the sole ionizing source in the western regions of the NLR. 
\end{itemize}

Overall, these results strongly suggest that photoionization \textit{and} collisional excitation commonly act as excitation mechanisms in the NLR of Mrk~3. This result is at odds with the canonical picture, which hypothesized that photoionization is the main excitation mechanism \citep{bianchi06,balmaverde12}. However, this canonical picture may be overly simplistic and does not reflect the complexity of Seyfert galaxies, most of which produce small-scale, weak, bipolar radio-emitting jets \citep{thean00,lal04}. Indeed, small-scale radio jets that are confined within the host galaxy are expected to interact with the surrounding dense interstellar material, which can give rise to shock heating \citep[e.g.][]{baum92,best00}. Therefore, it is feasible that shock heating plays a general, possibly complementing, role in the ionization of the gas surrounding the nuclei.

\begin{table}
\caption{Constraints on the gas outflow velocities}
\begin{minipage}{8.75cm}
\renewcommand{\arraystretch}{2}
\centering
\begin{tabular}{c | c c c c }
\hline 
Ion&  $-278$ pc$^\dagger$  & $-139$ pc$^\dagger$  & $139$ pc$^\dagger$  & $278$ pc$^\dagger$  \\
& $\rm{km \ s^{-1}}$ &$\rm{km \ s^{-1}}$ &$\rm{km \ s^{-1}}$ &$\rm{km \ s^{-1}}$ \\
\hline
Si XIV & $146^{+340}_{-146}$ & $97^{+340}_{-146}$ & $49^{+243}_{-194}$ & $291^{+631}_{-146}$ \\ 

Si XIII (R) & $-181^{+226}_{-181}$ & $181^{+271}_{-451}$ & $361^{+496}_{-496}$ & $857^{+722}_{-948}$ \\ 

Si XIII (F) & $-45^{+223}_{-579}$ & $45^{+1024}_{-312}$ & ... & ... \\ 

Mg XII & $36^{+428}_{-321}$ & $71^{+356}_{-321}$ & $35^{+107}_{-107}$ & $285^{+321}_{-107}$ \\ 

Ne X & $74^{+247}_{-297}$ & $74^{+247}_{-296}$ & $0^{+247}_{-445}$ & $-25^{+371}_{-272}$ \\ 
\hline

\end{tabular} 
\flushleft
$^\dagger$ Negative and positive offsets correspond to east and west, respectively. \\
$^\ddagger$(R) and (F) represent the resonance and forbidden lines, respectively.  \\
\end{minipage}
\vspace{0.75cm}
\label{tab:list3}
\end{table}

\subsection{Large-scale gas}
\label{sec:large}

To study the diffuse emission on galaxy scales, we extract an X-ray energy spectrum using an elliptical region with $54.5\arcsec$ and  $38.2\arcsec$ axis radii with position angle of $20\degr$. This region corresponds to the total elliptical aperture radius of the galaxy as measured by the 2MASS Large Galaxy Atlas \citep{jarrett03}. Since we aim to study the large-scale diffuse emission, we omit the counts originating from the NLR by excluding en elliptical region with $3.3\arcsec \times 2.2\arcsec$ radii centered on the center of Mrk~3.

To fit the spectrum of the large-scale diffuse emission, we employ a two component model consisting of an absorbed \textsc{apec} thermal emission model and a power law model. As before, we fixed the column density at the Galactic value and the slope of the power law at $\Gamma=1.56$.  We find a best-fit temperature and abundance of $kT=0.77\pm0.05$ keV and $Z=0.09^{+0.08}_{-0.04}$ Solar. With these parameters we obtain the absorption corrected $0.3-2$ keV band luminosity of $L_{\rm{0.3-2keV}} = 4.9\times10^{40} \ \rm{erg \ s^{-1}}$, which corresponds to the bolometric luminosity of $L_{\rm{bol}} = 8.3 \times10^{40} \ \rm{erg \ s^{-1}}$. Based on the normalization of the spectrum, we compute the emission measure of the gas and compute the total gas mass following Section \ref{sec:basic} and obtain $M_{\rm gas} = 1.0\times10^{9} \ \rm{M_{\odot}}$. 

To place the X-ray luminosity and gas mass of the galaxy into a broader context, we compute the X-ray luminosity per unit K-band luminosity. We derive the K-band luminosity of the the galaxy based on its apparent K-band magnitude ($m_{\rm K} = 8.97$) and obtain $L_{\rm K} = 1.8\times10^{11} \ \rm{L_{\rm \odot}}$. Using the $0.3-2$ keV band X-ray luminosity, we find that the specific X-ray emissivity of Mrk~3 is $L_{\rm{0.3-2keV}}/M_{\rm \star} = 2.7\times10^{29} \ \rm{erg \ s^{-1} \ L^{-1}_{K,\odot}}$. This value exceeds that obtained in low luminosity ellipticals, but are comparable to emissivities found in more massive (non-BCG) ellipticals \citep[e.g.][]{bogdan11,goulding16}. 

Although the NLR demonstrated an outflow in the east-west direction, it is not clear whether the gas is expelled from the galaxy or it is retained in the gravitational potential well. If a galactic-scale outflow is present, it may be either powered by the the energy input of Type Ia Supernovae or from the AGN. In this picture, the outflowing gas is replenished by the stellar yields originating from evolved stars, which are estimated to shed mass at a rate of $0.0021 \ \rm{L_{K}/L_{K,\odot} \ M_{\odot} Gyr^{-1}} $ \citep{knapp92}. Given the K-band luminosity of Mrk~3, we estimate that the mass loss rate from evolved stars is $\dot{M} = 0.38 \ \rm{M_{\odot} \ yr^{-1}}$. This implies that the replenishment time scale of the total observed gas mass is about $t_{\rm{repl}} = 2.6\times10^9$ years. To lift the gas from the potential well of the galaxy, we require $E_{\rm lift} = 7.2 \dot{M} \sigma^2$ \citep{david06}, where $\sigma=274 \ \rm{km \ s^{-1}}$ corresponds to the central stellar velocity dispersion. We thus find that the total energy required to lift the gas is  $E_{\rm lift} = 4.1\times10^{41} \ \rm{erg \ s^{-1}}$. 

The available energy from Type Ia Supernova can be computed by assuming that each supernova releases $10^{51} \ \rm{erg \ s^{-1}}$ energy and by computing the Type Ia Supernova rate of the galaxy using the frequency established by \citet{mannucci05} and the  K-band luminosity of the galaxy. Hence we obtain the Type Ia Supernova frequency of $6.4\times10^{-3} \ \rm{yr^{-1}} $, implying the total energy of $E_{\rm{SNIa}} = 2.0\times10^{41} \ \rm{erg \ s^{-1}} $. This value falls factor of about two short of the energy required to lift the gas from the potential of Mrk~3, hinting that Type Ia Supernovae cannot provide sufficient energy to drive a galaxy-scale outflow. 

The minimum energy required to drive a galactic-scale outflow is about factor of five lower than the kinetic energy ($E_{\rm kin} \gtrsim 2\times 10^{42} \ \rm{erg \ s^{-1}}$) from the AGN \citep{capetti99}, hinting that the AGN is able to expel the gas from the galaxy. However, the large hot gas mass in the galaxy combined with the long replenishment time of the gas argues against the existence of a large-scale outflow that would remove the gas from the gravitational potential well of the galaxy. Instead, it is more likely that the energy from the AGN plays a role in heating the X-ray gas, possibly driving it to larger radii.

\section{Summary}
In this work we analyzed \textit{Chandra} X-ray observations of the NLR of Markarian 3. By combining imaging and grating spectroscopy data, we achieved the following conclusions: 
\begin{itemize}
\item We confirmed the presence of X-ray emitting gas in the NLR of the galaxy. The average gas temperature and metallicity is $kT=0.85$ keV and $Z=0.24$ Solar. 
\item We deconvolved the X-ray image to probe the structure of the gas at small angular scales. The X-ray morphology of the hot gas was confronted with the radio and [O~III] morphology. We found that while the X-ray gas exhibits an S-shaped morphology, which is similar to those observed in other wavelengths, the hot gaseous emission has a broader distribution than the radio or [O~III] emission. 
\item We demonstrated the presence of shocks towards the west ($M=2.4\pm0.3$) and towards the east ($M=1.5^{+1.0}_{-0.5}$). This detection suggests that shock heating due to the interaction between the radio jets and the dense interstellar material may play a non-negligible role in the ionization of the gas. 
\item Spectroscopic analysis of the Si XIII triplet (resonance, intercombination, forbidden) lines suggests that both photoionization and collisional ionization may excite the hot gas. 
\item Using the high-resolution spectra we compared the best-fit line centroids between the east and west sides of the NLR. We did not find statistically significant differences, which hints at low projected outflow velocities that are significantly lower than those inferred from the Rankine-Hugonoit jump conditions. This difference implies that the outflow likely propagates along the plane of the sky. 
\item Given the common nature of small-scale radio jets in Seyfert galaxies, it is feasible that collisional ionization plays a role in the excitation of the hot gas in the NLR of other Seyfert galaxies as well.

\end{itemize}

\smallskip

\begin{small}
\noindent
\textit{Acknowledgements.} We thank the referee for the constructive comments. This research has made use of \textit{Chandra}  data provided by the Chandra X-ray Center. The publication makes use of software provided by the Chandra X-ray Center (CXC) in the application package CIAO. In this work the NASA/IPAC Extragalactic Database (NED) has been used. We acknowledge the usage of the HyperLeda database (http://leda.univ-lyon1.fr). \'A.B., R.P.K, W.R.R acknowledges support for the Smithsonian Institution. F.A-S. acknowledges support from \textit{Chandra} grant GO3-14131X.
\end{small}

\newpage

Appendix A:

\begin{figure}[!b]
  \begin{center}
    \leavevmode
      \epsfxsize=7cm\epsfbox{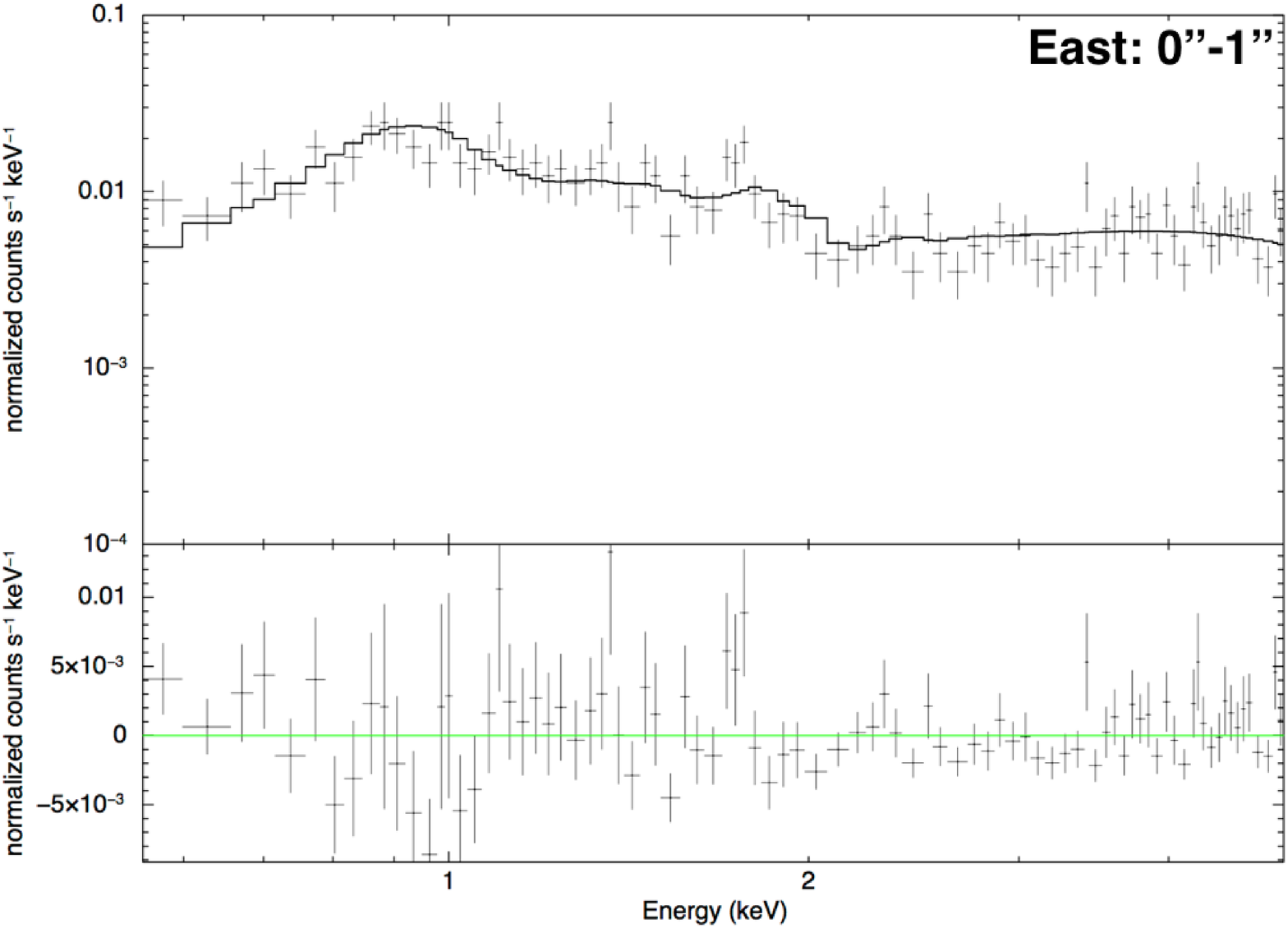}
      \epsfxsize=7cm\epsfbox{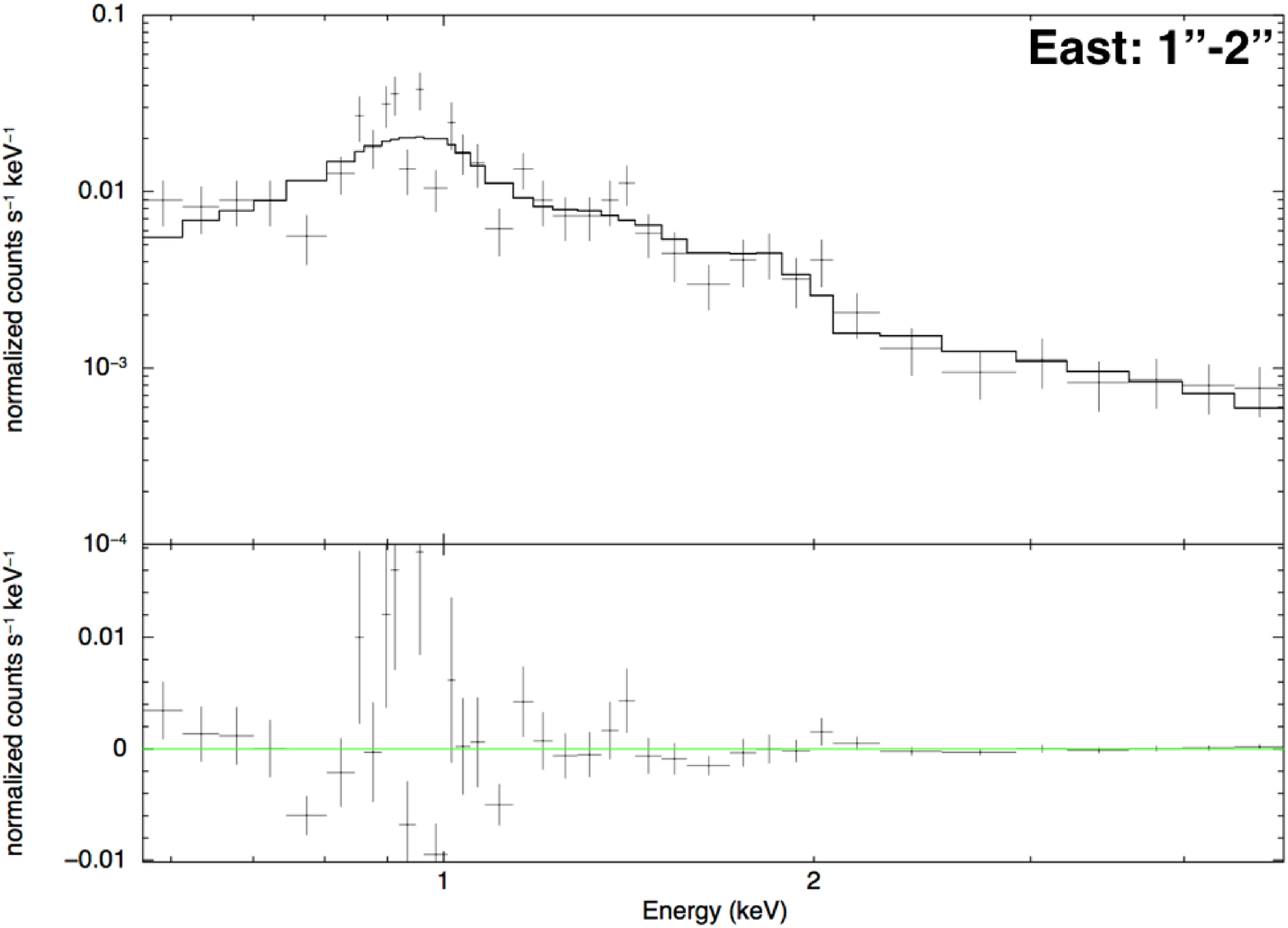}
      \epsfxsize=7cm\epsfbox{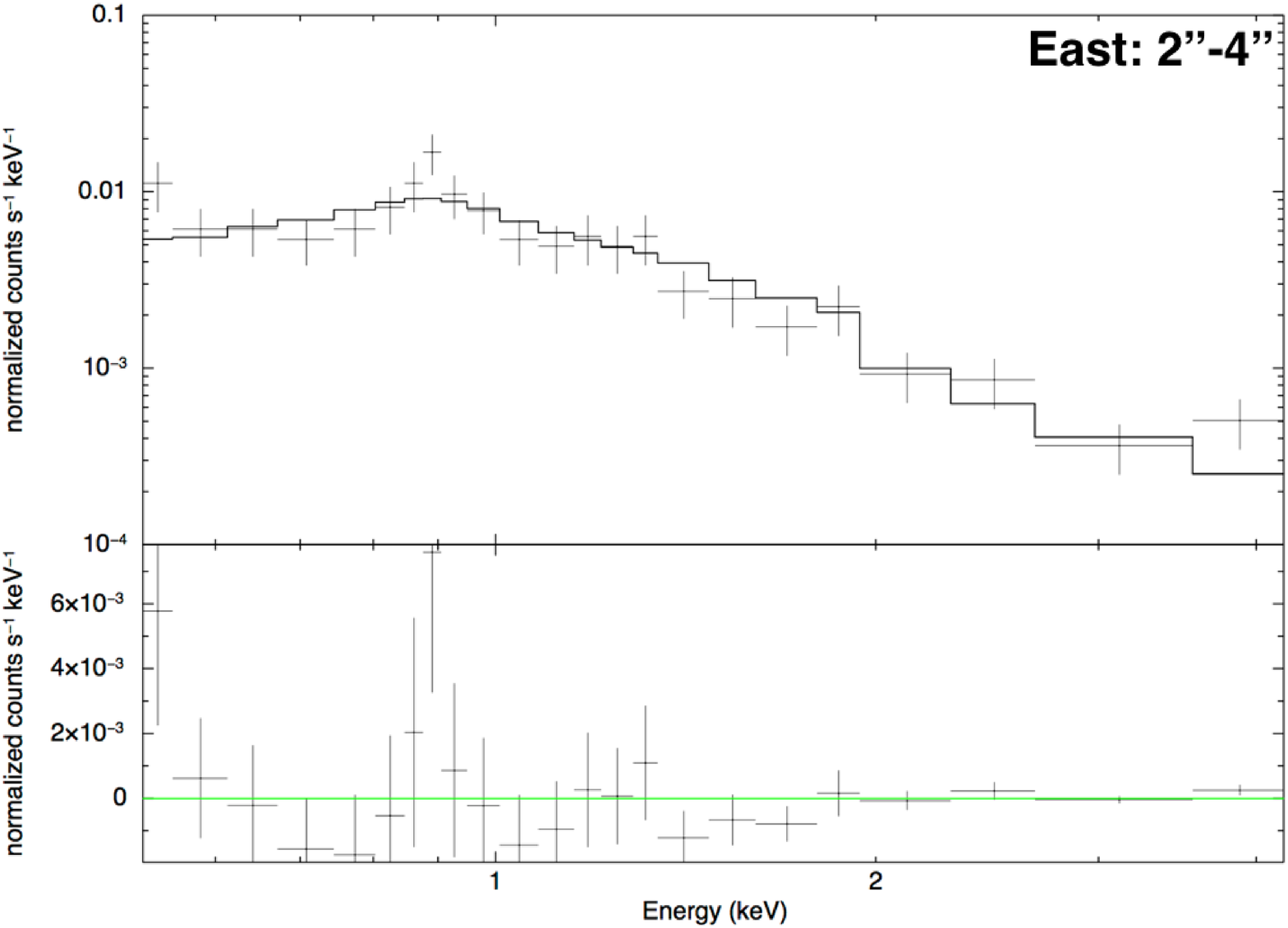}
      \epsfxsize=7cm\epsfbox{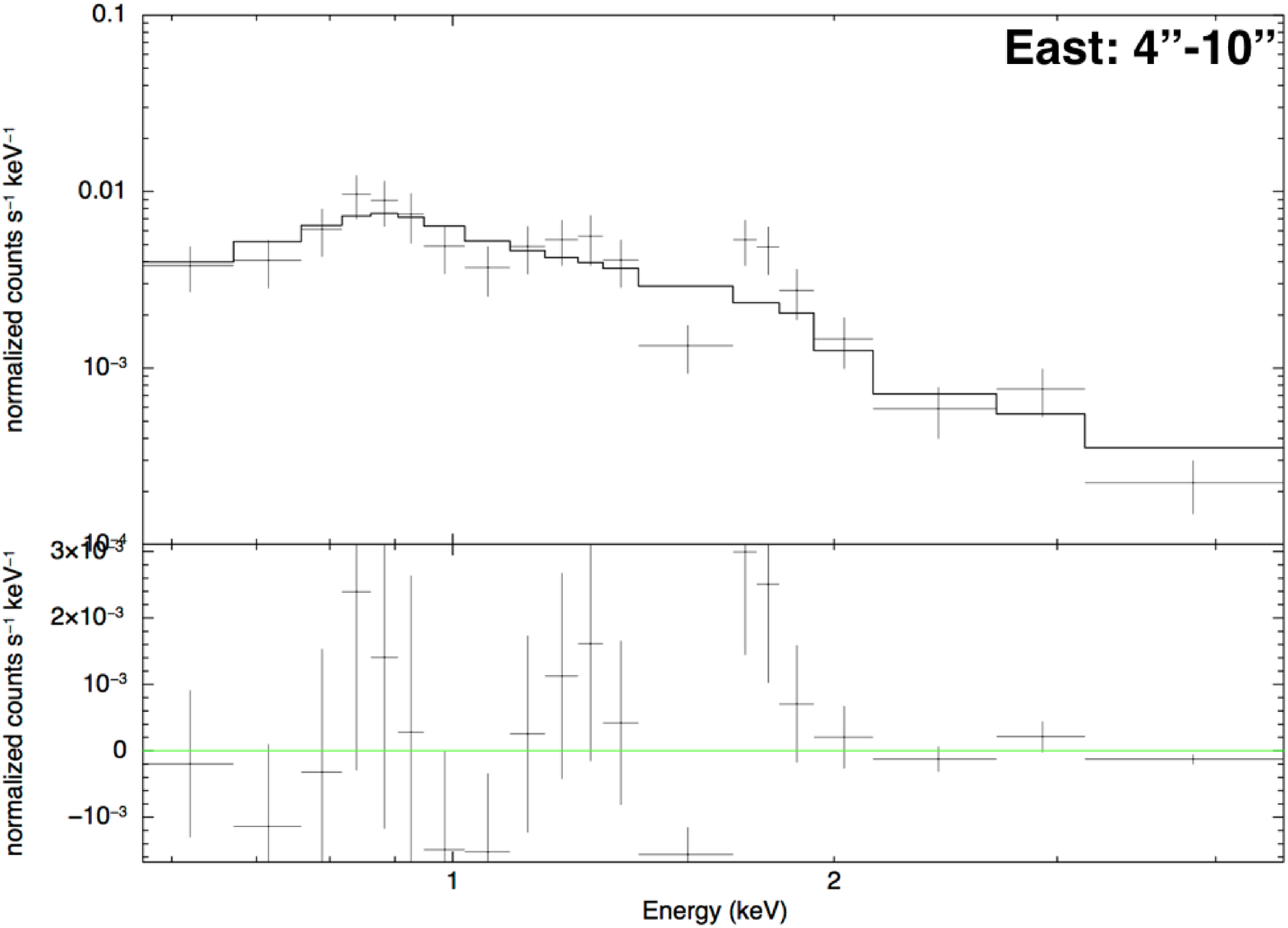}
      \epsfxsize=7cm\epsfbox{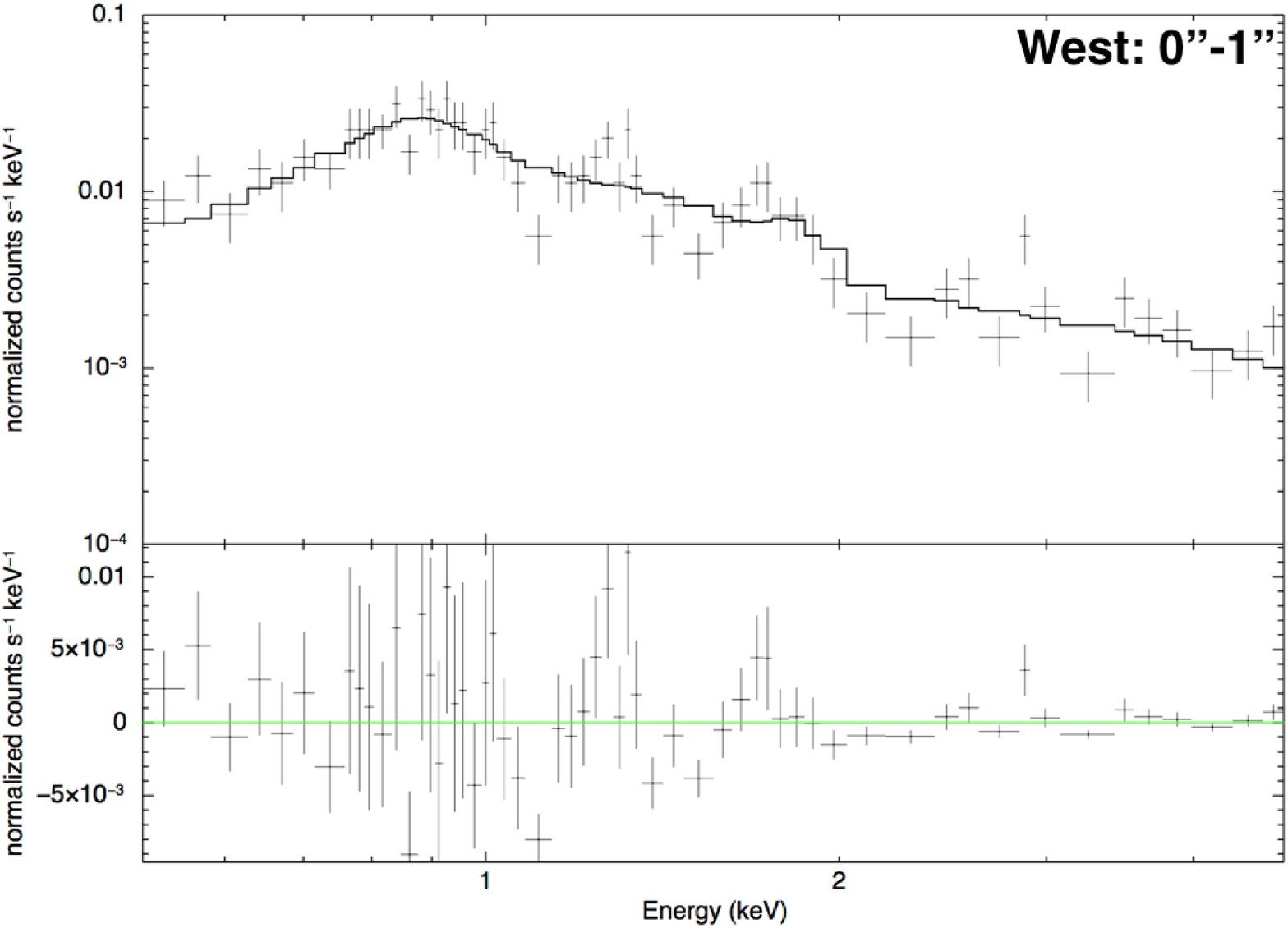}
      \epsfxsize=7cm\epsfbox{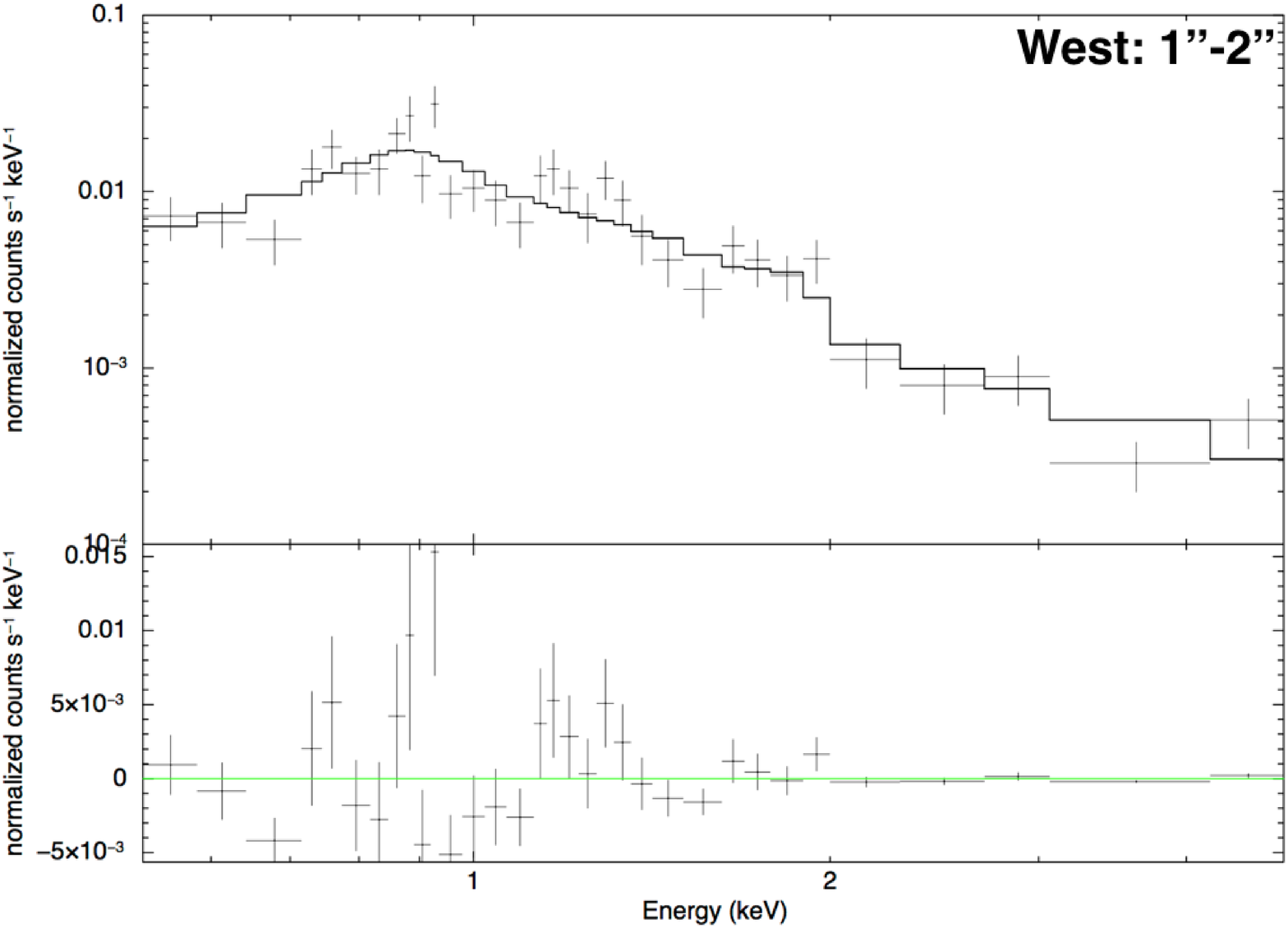}
      \epsfxsize=7cm\epsfbox{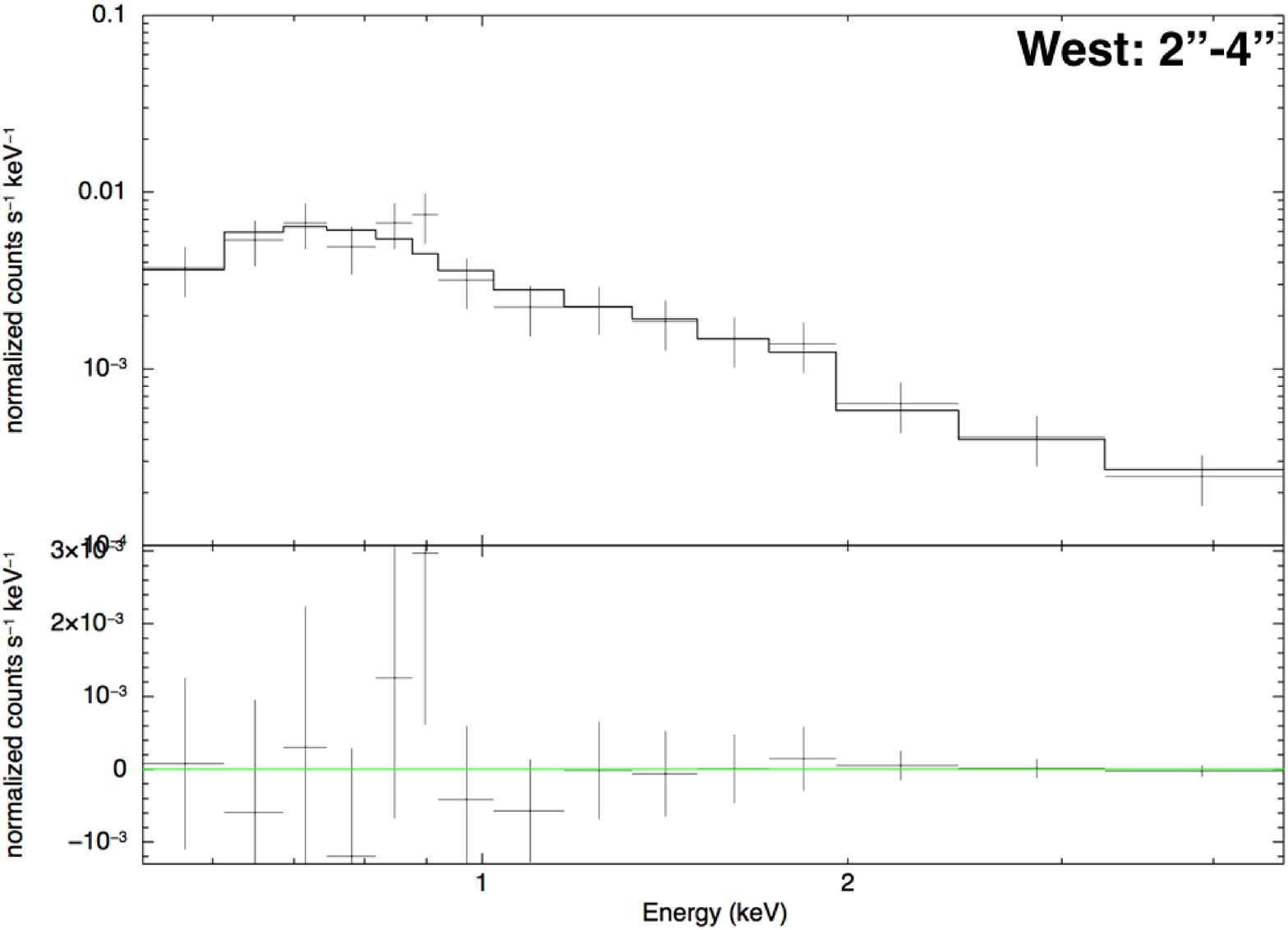}
      \epsfxsize=7cm\epsfbox{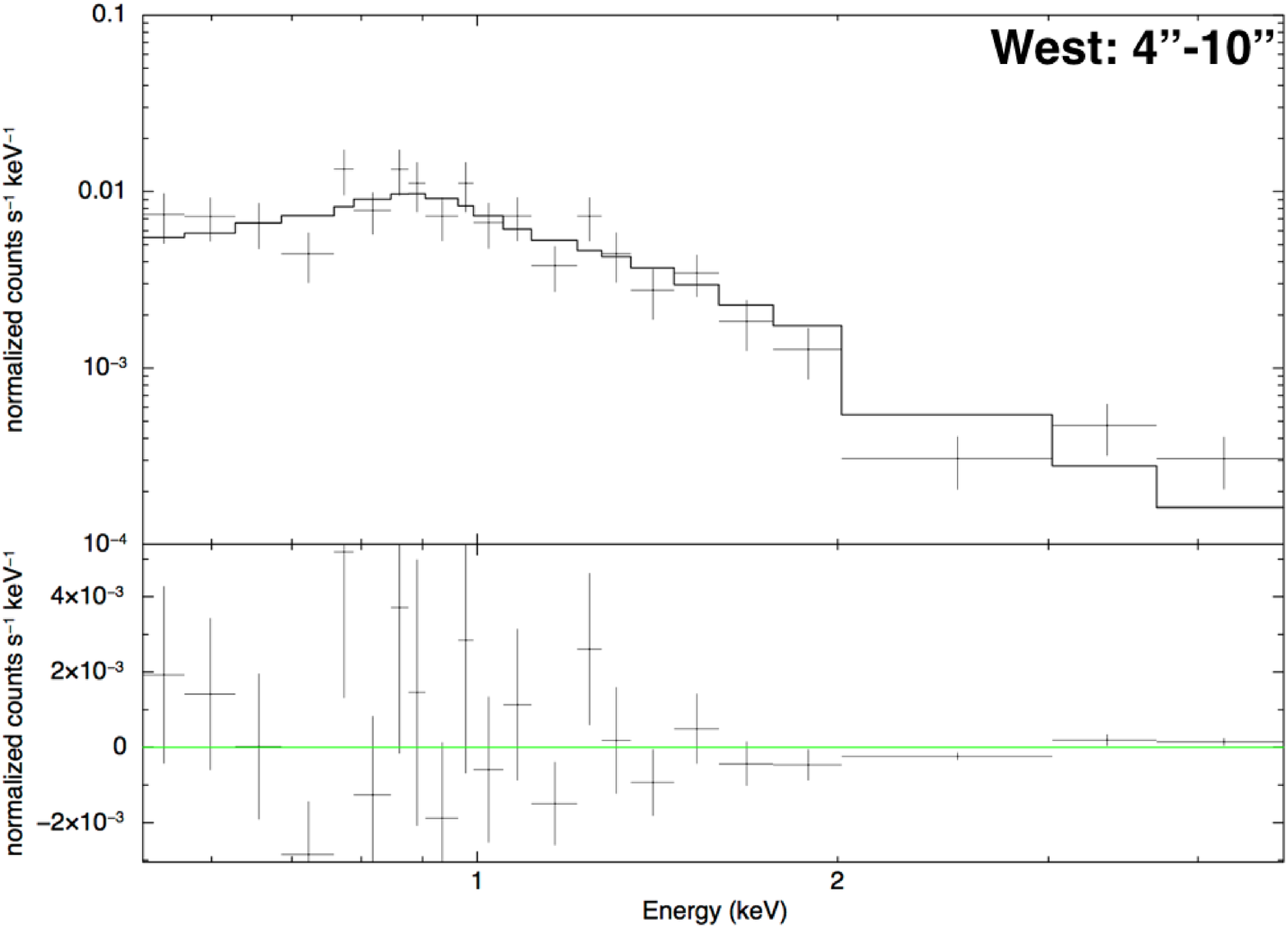}      
      \vspace{0cm}
      \caption{The X-ray energy spectra extracted towards the east and west of Mrk~3. These spectra were used to construct the temperature map, depicted in the right panel of Figure \ref{fig:sb}. The position angles used to extract the spectra were $135\degr-225\degr$ (east) and $315\degr-405\degr$ (west). The spectra were fit with a two component model consisting of an absorbed thermal model and a power law model. For details on the fitting procedure, see Section \ref{sec:shock}. The best-fit models are over plotted on the observed data points.}
     \label{fig:spec1}
  \end{center}
\end{figure}

\end{document}